\documentclass[twocolumn,twoside]{revtex4}
\usepackage{graphicx}
\usepackage{fancyhdr}
\usepackage{amsmath}
\include{babarsym}

\begin{document}

\title{Measurement of hadronic cross-sections with the \babar~detector}

%

\author{Marcus Ebert\\
on behalf of the \babar~Collaboration}
\affiliation{University of Victoria, BC, Canada, V8N2X3}

\begin{abstract} 

A program of measuring the light hadrons production in exclusive $\epem\to$ hadrons processes is in place at \babar~with the aim to 
improve the calculation of the hadronic contribution to the muon $g-2$. We present the most recent results obtained by using the full data 
set of about $514 \invfb$ collected by the \babar~experiment at the PEP-II \epem collider at a center-of-mass energy of about $10.6 \gev$. 
In particular, we report the results on the channels $\epem\to\pip\pim 3\piz$, $\epem\to \pip\pim 2\piz\eta$, and its resonant sub-states. 
These final states are studied in a wide mass range,from threshold production up to about 4\gevcc.
In addition to the cross-sections, twelve \jpsi and $\psitwos$ branching fractions were measured with ten of them be a first-time measurement.
\end{abstract}

\maketitle

\thispagestyle{fancy}


\section{Introduction}
\begin{figure}
\includegraphics[width=0.5\linewidth]{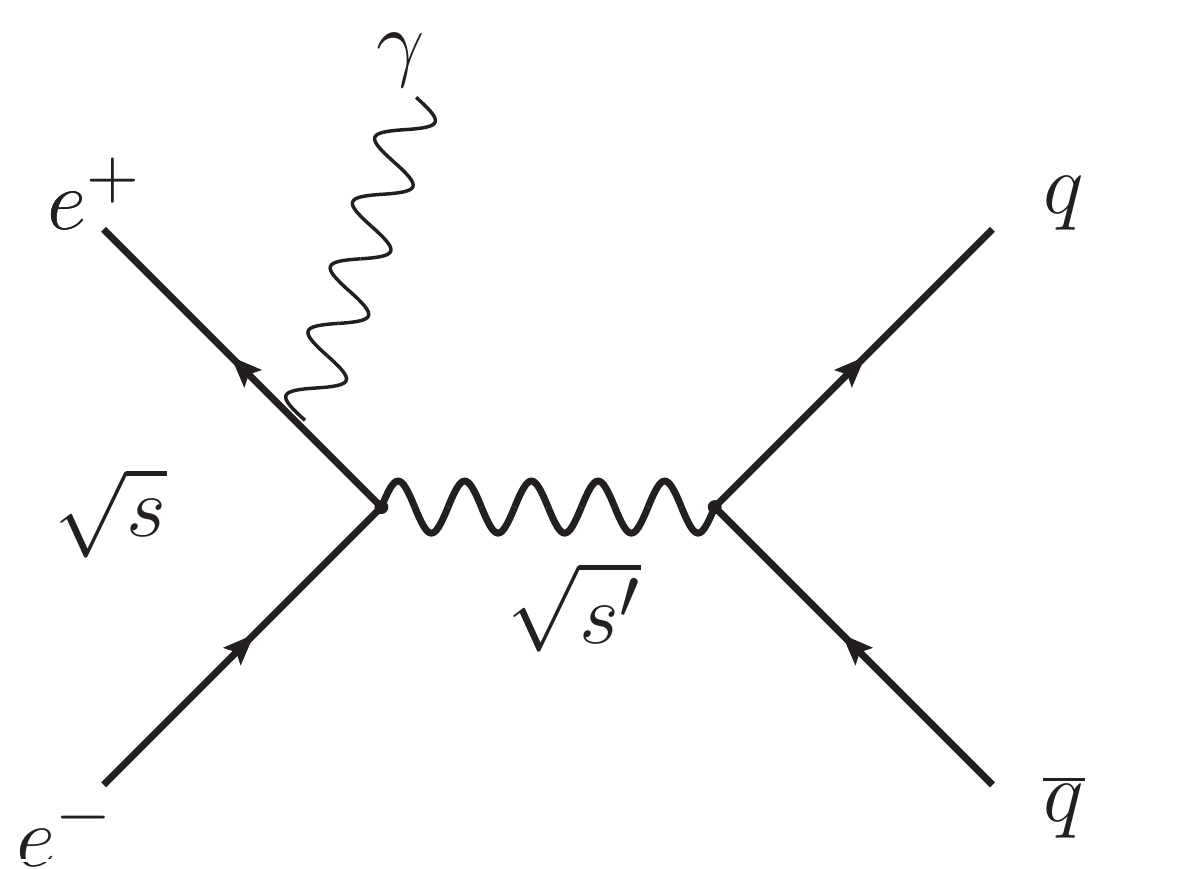}
\caption{Emitting a photon in the initial state reduces $\sqrt{s}$ to $\sqrt{s^\prime}$. \label{ebert:isrfig}}
\end{figure}

Study of the $(g_\mu-2)$ value show a discrepancy of about 3.7 standard deviation between the value measured in experiments and the value obtained from the Standard Model 
\cite{ebert:PDG}.
The calculation of the Standard Model value needs to account for hadronic vacuum polarization which is obtained from experimental \epem hadronic 
cross-section measurements, especially within low-energy regions. So far not all accessible states have been measured and future measurements will
improve the calculation of the Standard Model value for $(g_\mu-2)$. 
Using initial-state radiation (ISR) processes give access to a wide range of different center-of-mass (CM) energies. 
In ISR events, the beam-energy in the CM system, $\sqrt{s}$, is reduced to a new CM energy, $\sqrt{s^\prime}$, 
while the reduction depends on the value of the initial-state photon (fig.\ref{ebert:isrfig}).
In addition, ISR  events also allow resonance spectroscopy over a wide energy region.

\babar~is an \epem collider experiment which has a rich ISR program and measured many different hadronic cross-sections in ISR events over the years. 
During the data taking period 1999-2008, \babar~collected 514\invfb of data at CM energies equivalent to the mass of different $\bbbar$ resonances.
\babar~published more than 580 papers over the years, and even thought it has been 11 years since \babar~stopped taking data there are still many ongoing analyses.

In the analysis presented here, ISR events over an energy region from threshold up to the charmonium region are studied using the final states $\pip\pim 2\piz\eta$ 
and $\pip\pim 3\piz$, including its resonant sub-states.

\section{ISR study of $\pmb{\epem\to\pip\pim 3\piz}$ and $\pmb{\epem\to\pip\pim\piz\piz\eta}$}
\subsection{Previous results}
\begin{figure}
\includegraphics[width=0.7\linewidth]{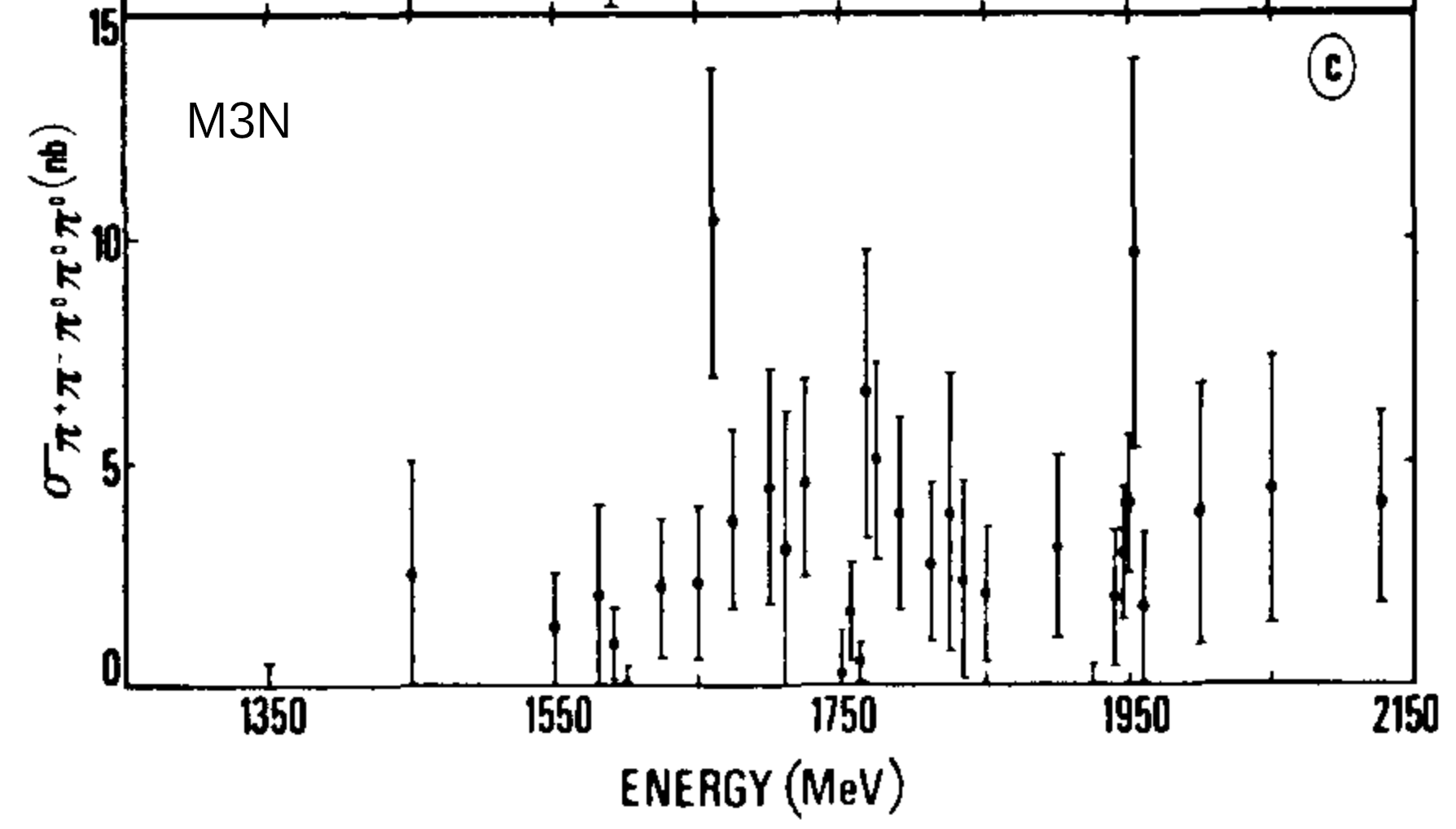}
\caption{Cross-section measurement of $\epem\to\pip\pim 3\piz$ by the M3N collaboration.\label{ebert:M3Nfig}}
\end{figure}

\begin{figure}
\includegraphics[width=0.9\linewidth]{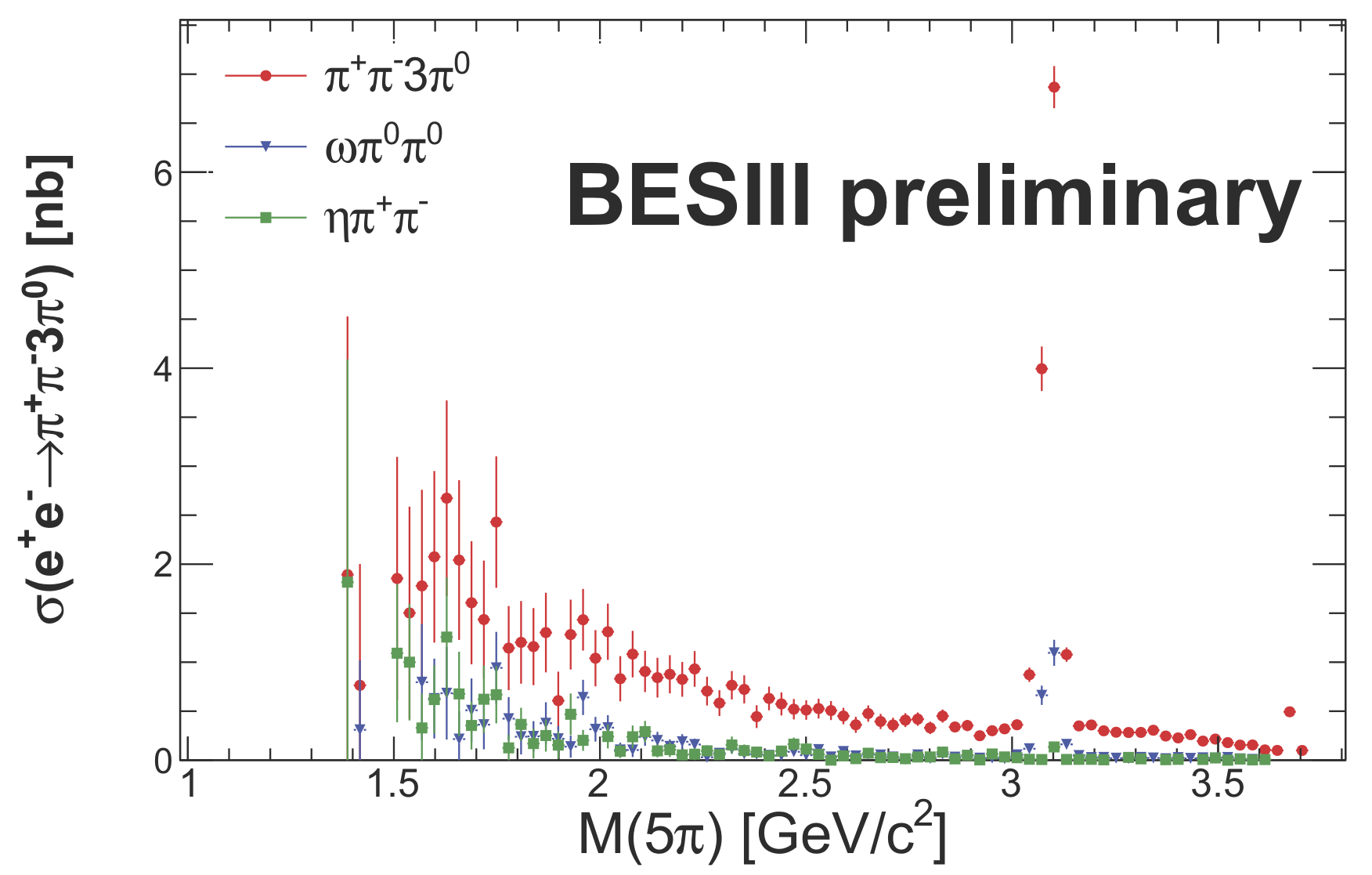}
\caption{Preliminary cross-section measurement of $\epem\to\pip\pim 3\piz$ by the BESIII collaboration. \label{ebert:besiiifig}}
\end{figure}

The cross-section for $\epem\to\pip\pim 3\piz$ was reported by M3N \cite{ebert:M3N} and MEA \cite{ebert:MEA} in 1979. While MEA reported only a single cross-section
value over the whole energy range they studied, the M3N result is shown in fig.\ref{ebert:M3Nfig}. Recently, BESIII also showed a preliminary
cross-section measurement for $\epem\to\pip\pim 3\piz$ \cite{ebert:besiii}, including the cross-section for the resonant sub-states (fig.\ref{ebert:besiiifig}).

\babar~already measured the cross-section for $\epem\to\eta\pip\pim$ with $\eta\to\pip\pim\piz$ and $\eta\to\gamma\gamma$. In this presentation, the cross-section measurement
for the decay mode $\epem\to\eta\pip\pim$ with $\eta\to 3\piz$ as a resonant sub-state of $\epem\to\pip\pim 3\piz$ is presented.

A measurement for the cross section of $\epem\to\pip\pim\piz\piz\eta$ has not been reported by another experiment so far. 
The SND Collaboration however reported cross-section 
measurements for the resonant sub-modes $\epem\to\omega\piz\eta$ and $\epem\to\phi\piz\eta$.

\subsection{Selection criteria}
For the selection of events of interest, two well reconstructed tracks are required. Those tracks are fitted to a common vertex which needs to be close
to the collision point of the beams. In addition, both tracks need to be inconsistent with being a kaon or muon. Also, each event needs to have at least seven
photons. At least one photon needs to have an energy of more than $3\gev$ and it is assumed that the photon with the highest energy is the ISR photon. 
The additional six photons are combined into three pairs resulting in fifteen combinations when all possible combinations are considered.
At least two of those pairs, the \piz candidates, need to have a mass that is within 35\mevcc around the \piz-mass, 
$\|m_{\piz_\textrm{cand}}-m_{\piz}\|<35\mevcc$. There is no constraint placed on the third pair which allows to reconstruct the final 
state with 3\piz as well as the final state with $2\piz\eta$. A kinematic fit is performed on the the whole event, $\epem\to\pip\pim 2\piz\gamma\gamma\gamma_\mathrm{ISR}$, with the following conditions:
\begin{itemize}
\item the mass of the \piz candidates, $m_{\piz_\textrm{cand}}$, is constrained to the mass of a \piz, $m_{\piz}$,
\item $\chi^2_{\pip\pim\piz\piz\gamma\gamma}$, obtained from the kinematic fit, needs to be smaller than 60, and
\item the combination of candidates that gives the smallest $\chi^2$ in this fit will be used.
\end{itemize}
In addition, a control region is defined as the region where  $\chi^2_{\pip\pim\piz\piz\gamma\gamma}$ is greater than 60 and smaller than 120.

\subsection{Background reduction}
The main background comes from $\tau^+\tau^-$ decays, from $\epem\to\pip\pim\piz\piz\gamma_\mathrm{ISR}$ events, and from non-ISR $uds$ events.
To suppress background from $\tau^+\tau^-$ decays, charged tracks are required to be not too close to the ISR photon. 

The cross-section for the ISR mode $\epem\to\pip\pim\piz\piz$ is expected to be larger than for the studied decay modes. 
When there are two additional background photons in the event, 
then it could appear to be a signal in one of the studied decay modes. 
To suppress those events, a kinematic fit is also performed for this final state and events with
$\chi^2_{\pip\pim\piz\piz\gamma_\mathrm{ISR}}<30$ are rejected.

The non-ISR event $\epem\to\pip\pim\piz\piz\piz\piz$ could also appear as a signal when one of the \piz decays in a way that one of the two photons is very high energetic. 
In this case, this high-energy photon would be mis-identified as the ISR-photon. Since the second photon has a very low energy in this case, those events would look like signal.
Therefore this kind of background is taken into account when determining in the cross-section.

\subsection{Signal extraction}
By varying the required ISR-photon energy in 50\mev steps, $\sqrt{s^\prime}$ is scanned from threshold up to the charmonium region. In each 50\mev interval, the signal is 
then extracted as described in this subsection.

\begin{figure}
\includegraphics[width=0.9\linewidth]{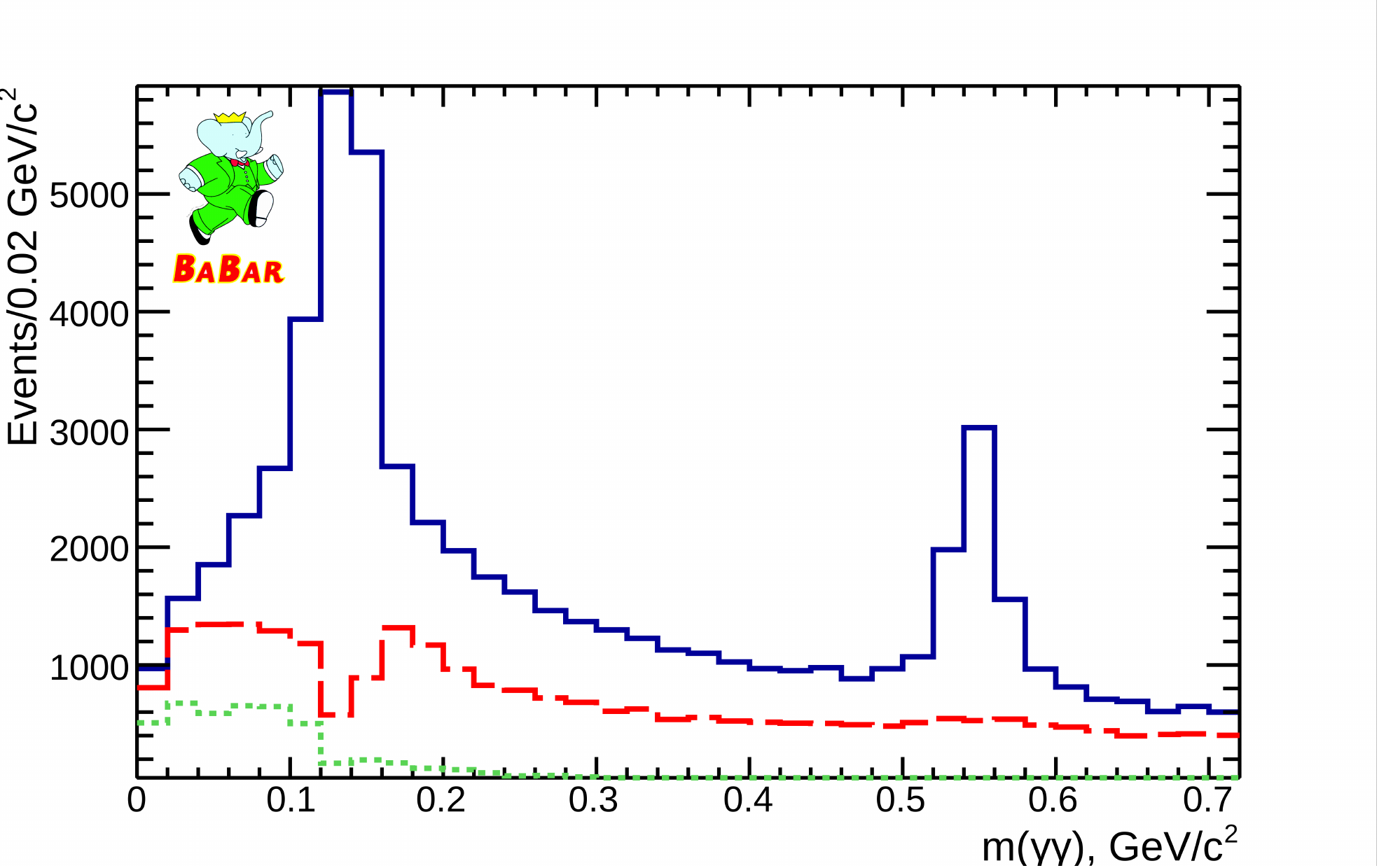}
\caption{Invariant $\gamma\gamma$ mass distribution in $\epem\to\pi\pim\piz\piz\gamma\gamma$ ISR events obtain in the signal region(blue) and the control region
 (red), as well as background events from $\epem\to\pip\pim\piz\piz$ (green).  \label{ebert:mgg}}
\end{figure}

\begin{figure}
\includegraphics[width=0.9\linewidth]{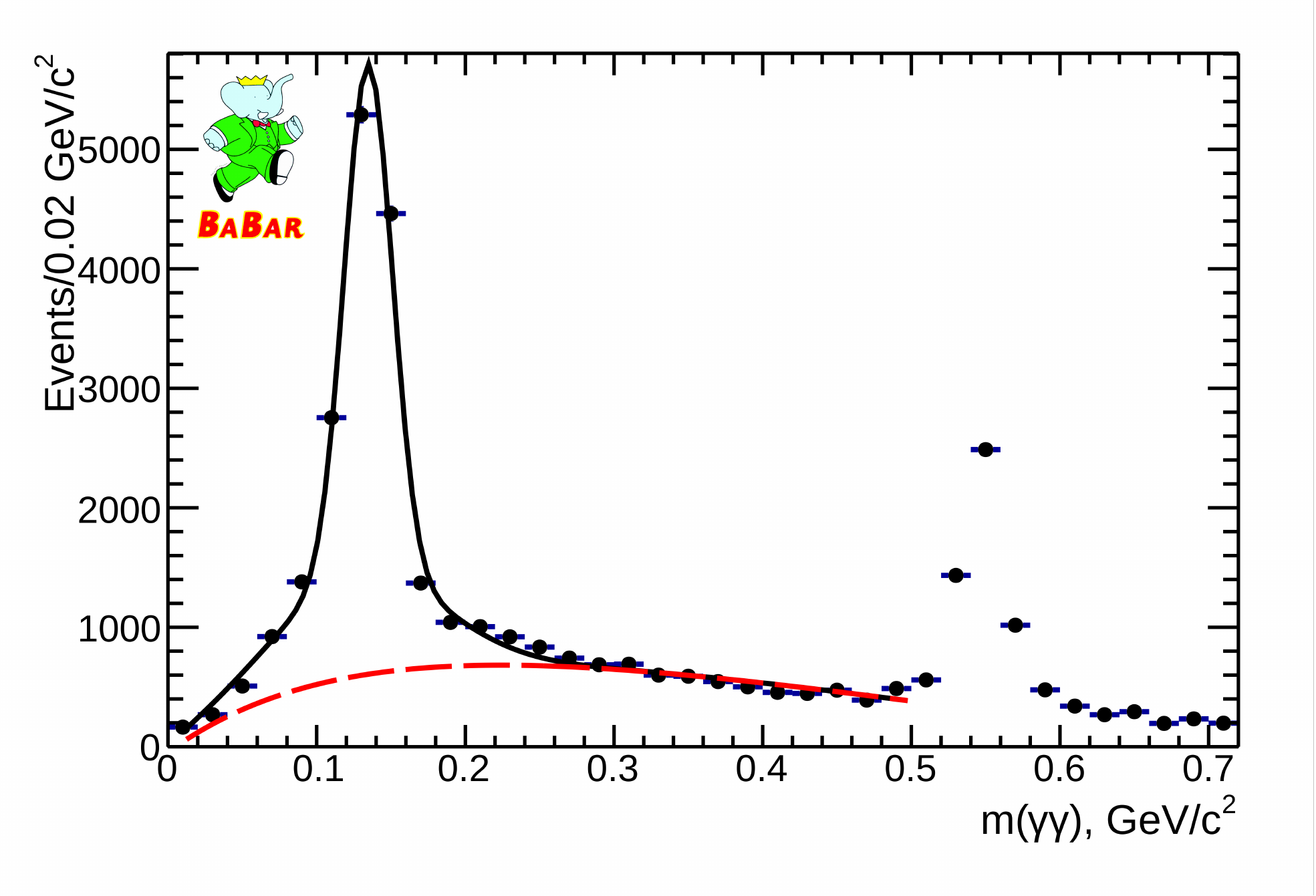}
\caption{$m(\gamma\gamma)$ distribution after subtracting the distribution from the control region (black markers) fitted for the number of \piz (black line) and
background (red). \label{ebert:mggsubtr}}
\end{figure}

The signal extraction for the modes $\epem\to\pip\pim 3\piz$ and $\epem\to\pip\pim 2\piz\eta$
is done in the invariant mass distribution of the third $\gamma\gamma$ pair. In each interval,
the $m(\gamma\gamma)$ distribution from the $\chi^2$ control region is subtracted from the signal distribution and the
resulting distribution fitted for the number of \piz or $\eta$, depending on the final state. In those fits, the signal shape is
fixed to the one obtained from simulated signal events. Figure \ref{ebert:mgg} shows the $m(\gamma\gamma)$ distribution for all energy intervals combined and 
fig.\ref{ebert:mggsubtr} the fit to obtain the number of \piz after subtracting the control region distribution.


When studying resonant sub-modes, the same procedure is used but the number of signal events in each interval is obtained from a fit to the invariant mass
distribution that shows the resonance. The obtained signal yields are then used to determine the cross-sections. 

\section{Results}
\subsection{$\pmb{\pip\pim 3\piz}$ and $\pmb{\pip\pim 2\piz\eta}$ final states}
\begin{figure}
\includegraphics[width=0.9\linewidth]{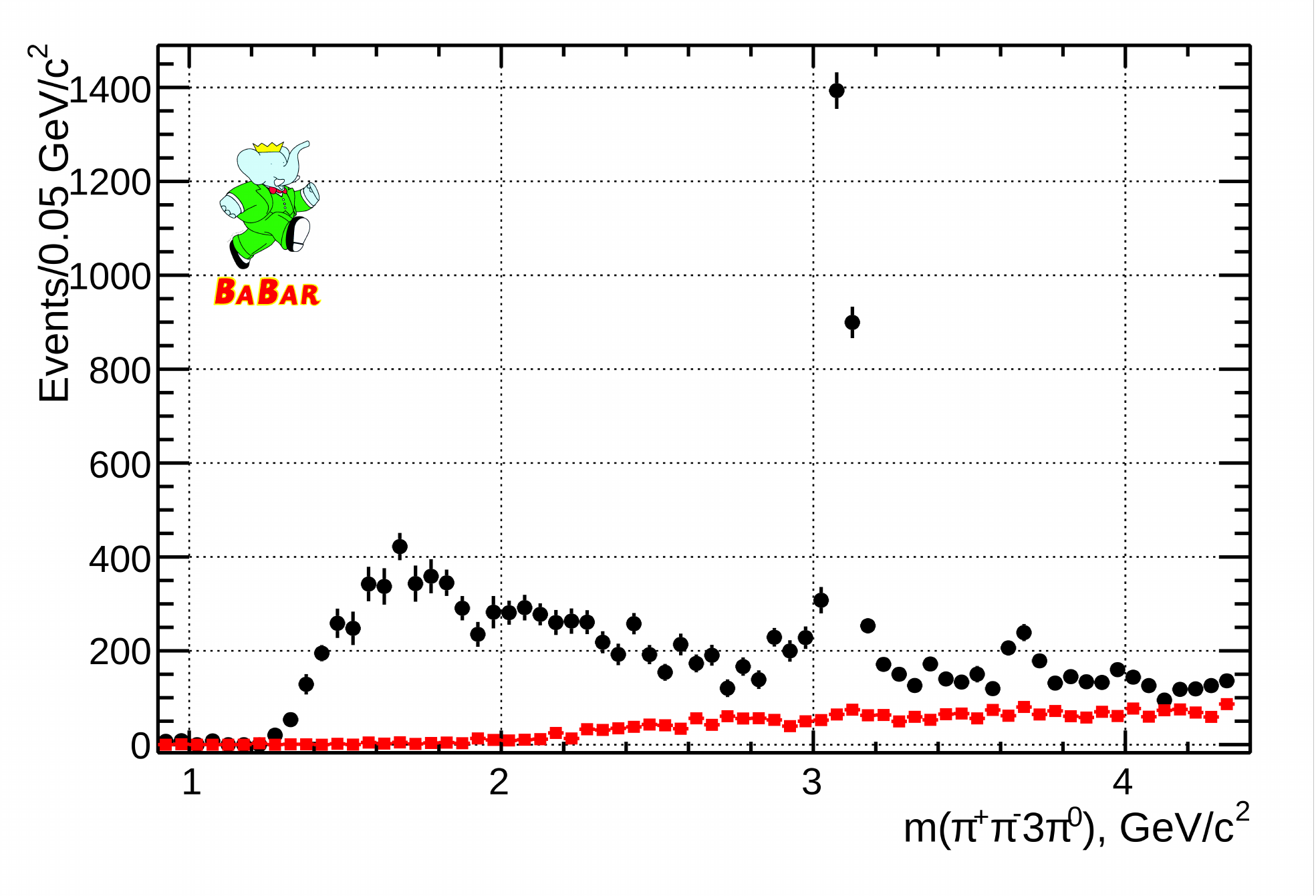}
\caption{Invariant mass distribution of $\epem\to\pip\pim 3\piz$ ISR events (black) and background from non-ISR events (red). \label{ebert:5pieventsfig}}
\end{figure}

\begin{figure}
\includegraphics[width=0.9\linewidth]{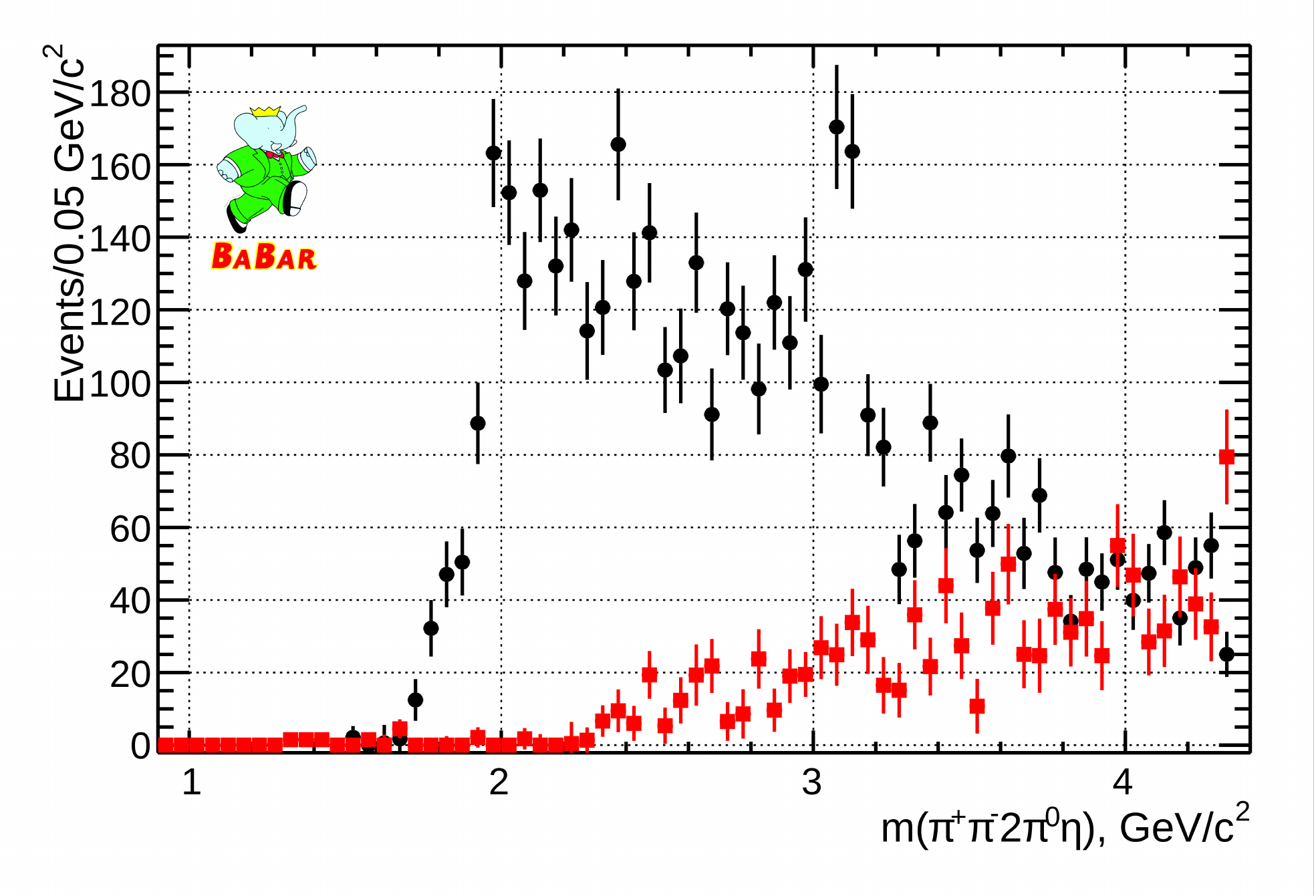}
\caption{Invariant mass distribution of $\epem\to\pip\pim 2\piz\eta$ ISR events (black) and background from non-ISR events (red). \label{ebert:2pietaeventsfig}}
\end{figure}

\begin{figure}
\includegraphics[width=0.9\linewidth]{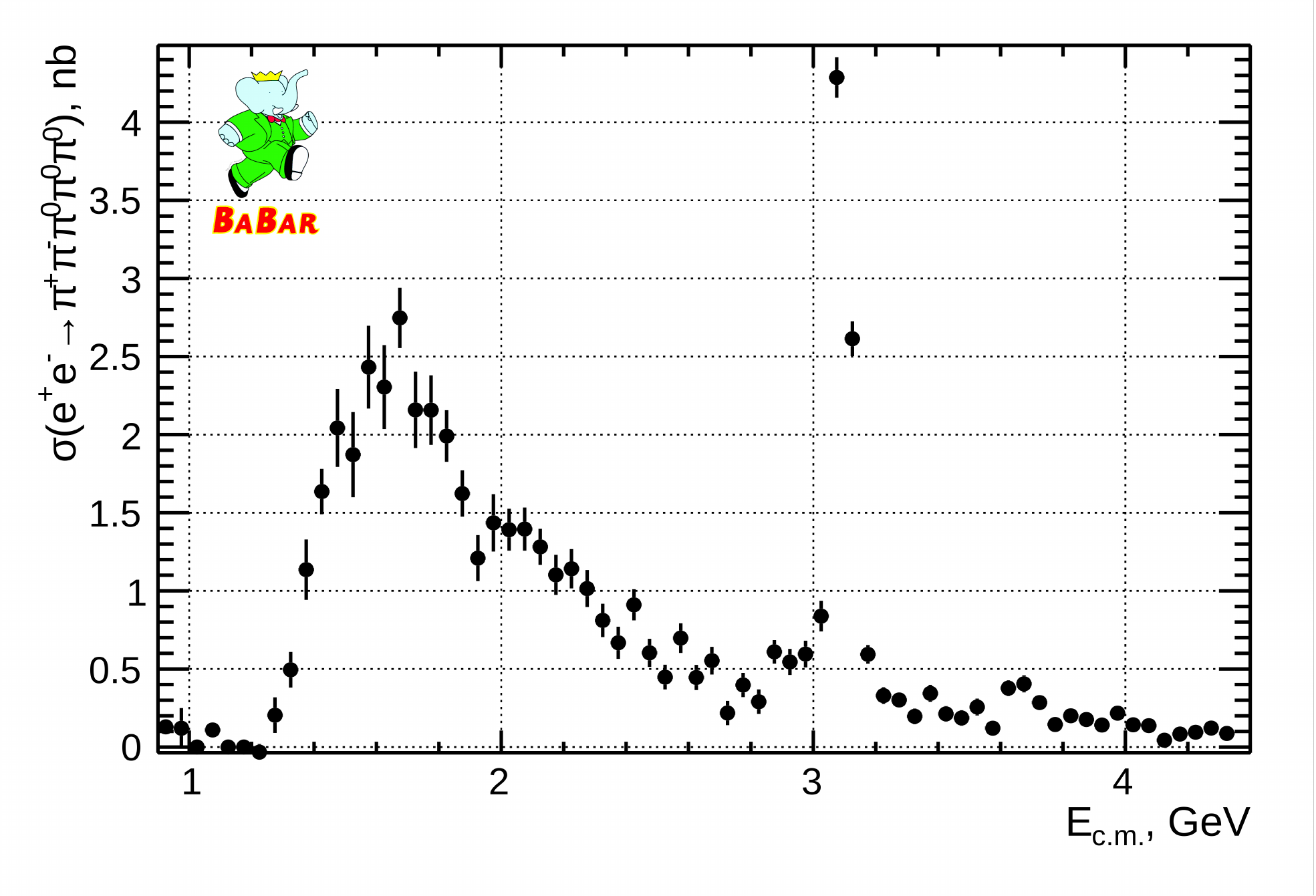}
\caption{Cross-section for $\epem\to\pip\pim 3\piz$ ISR events in 50\mev intervals in $\sqrt{s^\prime}$ ($E_\mathrm{c.m.^\prime}$). \label{ebert:5picrosssectionfig}}
\end{figure}

\begin{figure}
\includegraphics[width=0.9\linewidth]{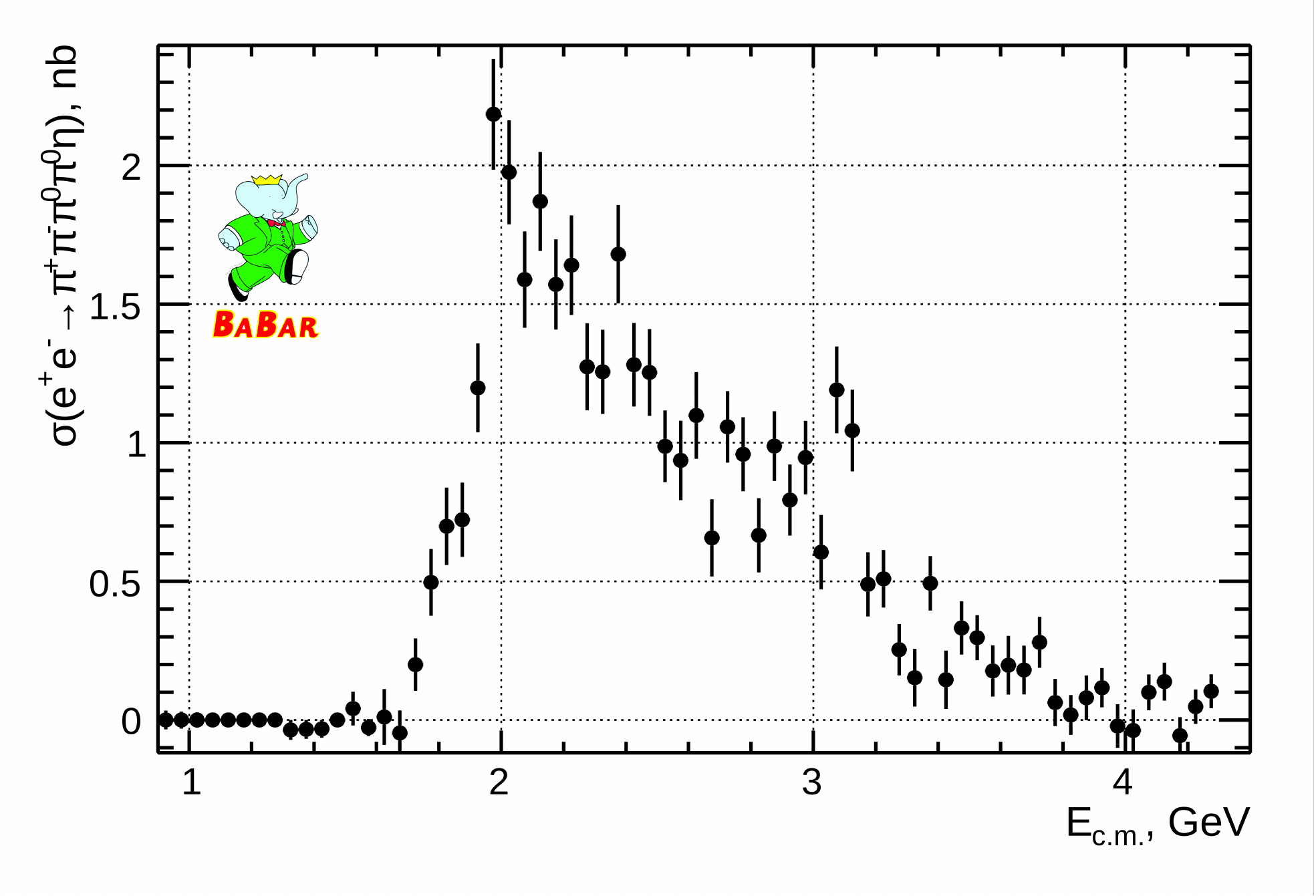}
\caption{Cross-section for $\epem\to\pip\pim 2\piz\eta$ ISR events in 50\mev intervals in $\sqrt{s^\prime}$ ($E_\mathrm{c.m.^\prime}$). \label{ebert:2pietacrosssectionfig}}
\end{figure}

Using the fit results for the number of signal events from each $\sqrt{s^\prime}$ interval, a 
signal distribution over the whole accessible energy range is obtained for each final state. 
Those distributions are shown for the $\pip\pim 3\piz$ final state in fig.\ref{ebert:5pieventsfig} 
and for the $\pip\pim 2\piz\eta$ final state in fig.\ref{ebert:2pietaeventsfig}.
Those distributions are then used to determine the cross-section for each interval, also taking the additional background contributions into account. 
The resulting cross-section distributions are shown in fig.\ref{ebert:5picrosssectionfig} and in fig.\ref{ebert:2pietacrosssectionfig}.

\subsection{$\pmb{\pip\pim\eta}$ and $\pmb{\omega\piz\piz}$ final states}
\begin{figure}
\includegraphics[width=0.9\linewidth]{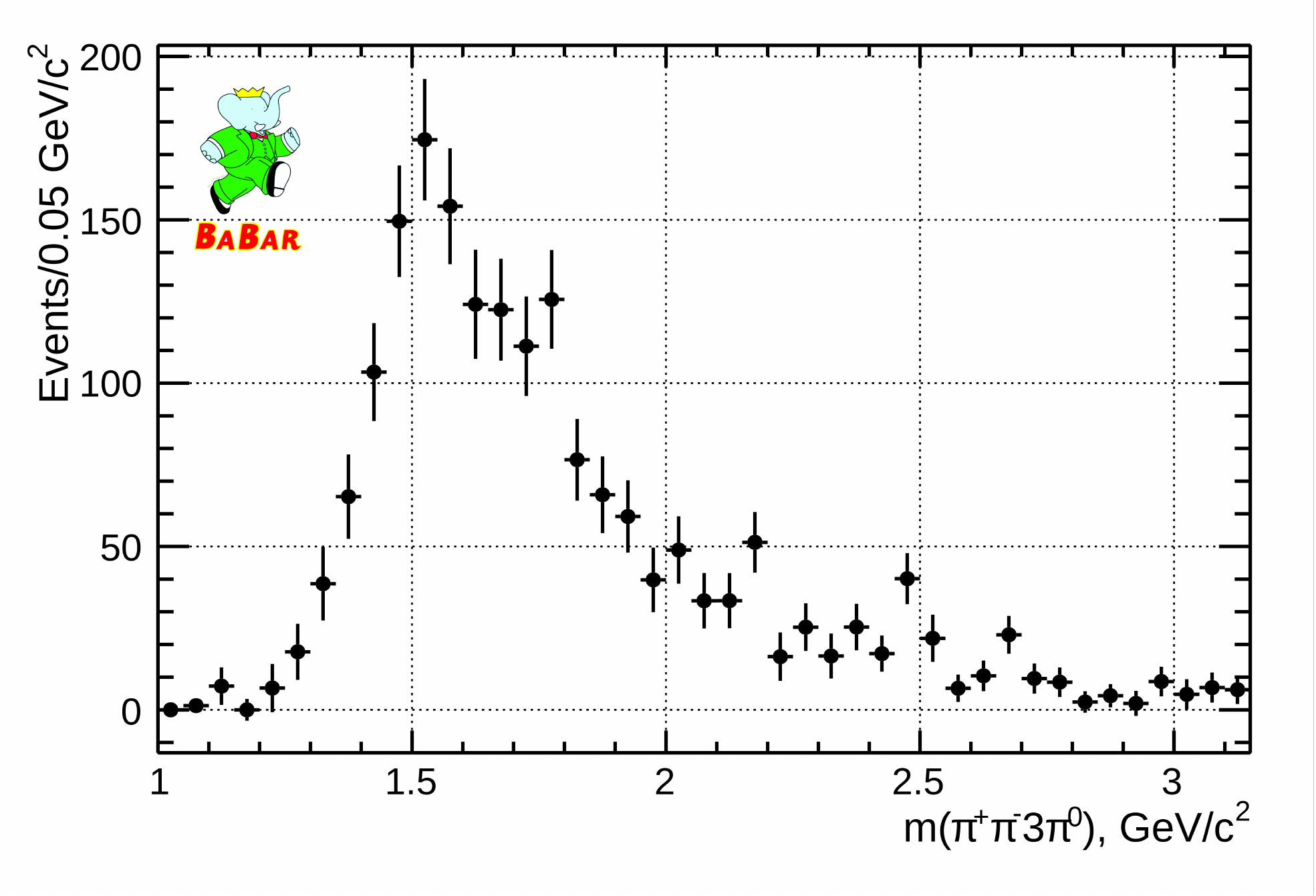}
\caption{Invariant mass distribution $m(\pip\pim 3\piz)$ for $\epem\to\pip\pim \eta, \eta\to 3\piz$ ISR events. \label{ebert:etaeventsfig}}
\end{figure}

\begin{figure}
\includegraphics[width=0.9\linewidth]{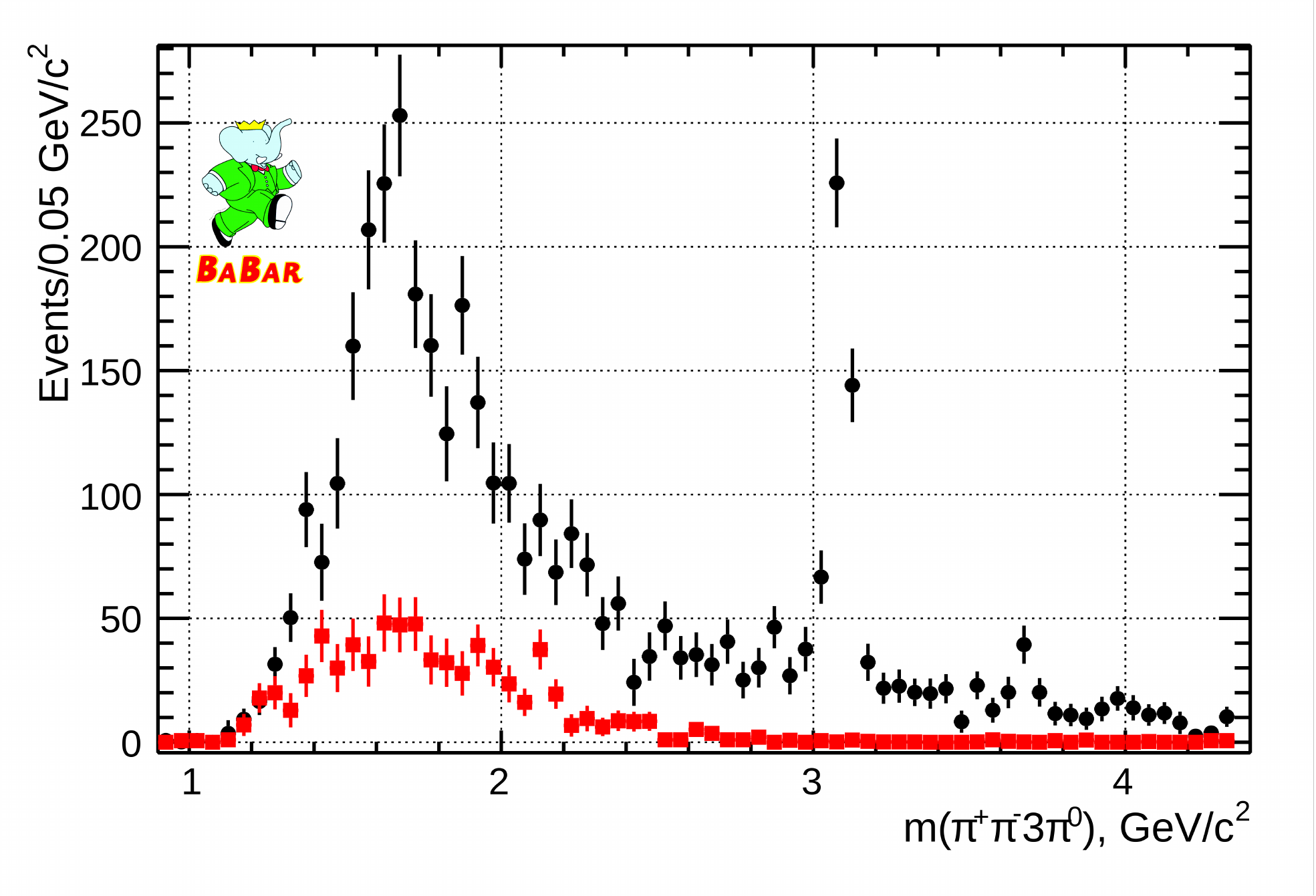}
\caption{Invariant mass distribution $m(\pip\pim 3\piz)$ for $\epem\to\omega\piz\piz, \omega\to\pip\pim\piz$ ISR events (black) and background from $\epem\to\omega\piz$ events.
 \label{ebert:omegaeventsfig}}
\end{figure}

\begin{figure}
\includegraphics[width=0.9\linewidth]{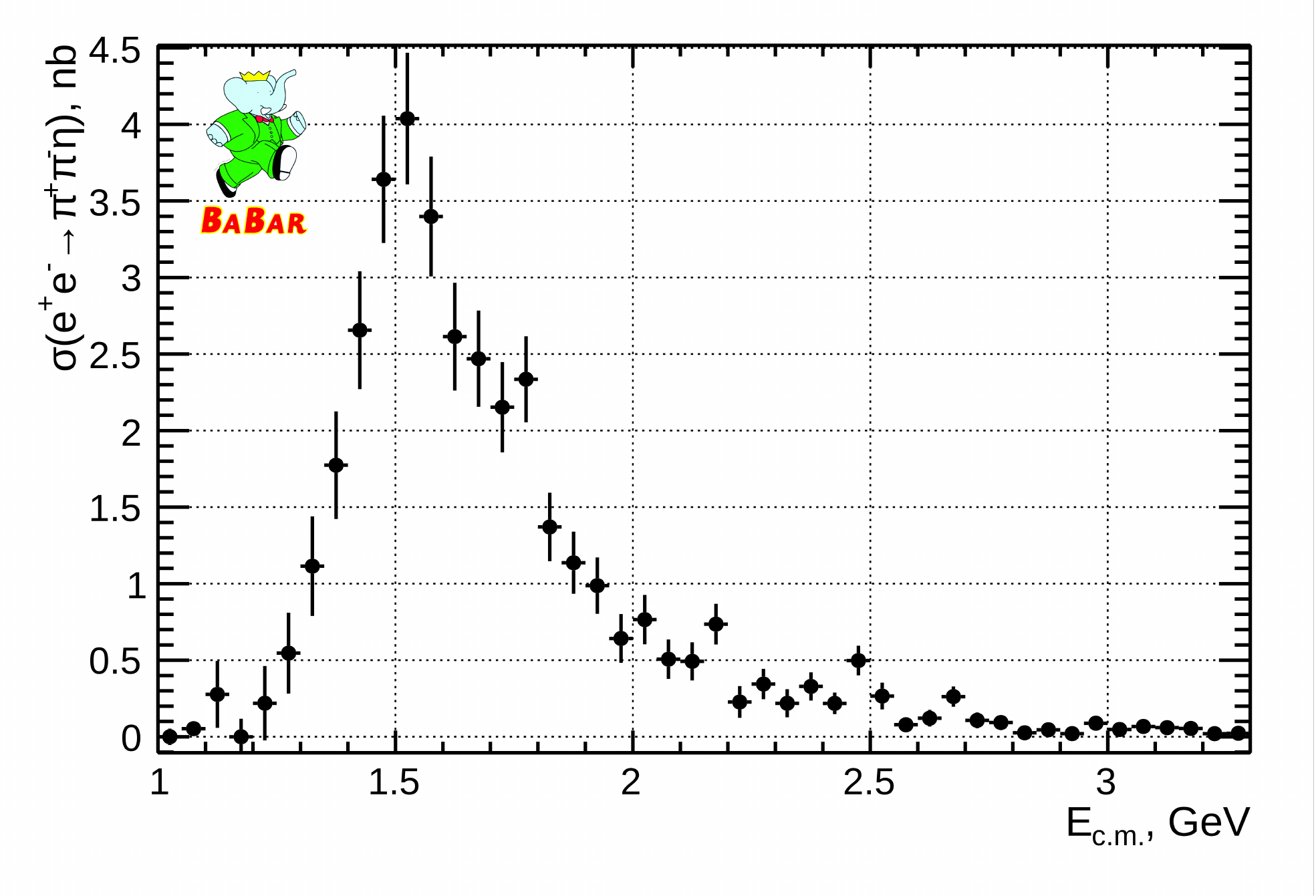}
\caption{Cross-section for $\epem\to\pip\pim\eta$ ISR events in 50\mev intervals in $\sqrt{s^\prime}$ ($E_\mathrm{c.m.^\prime}$). \label{ebert:xsectionetafig}}
\end{figure}

\begin{figure}
\includegraphics[width=0.9\linewidth]{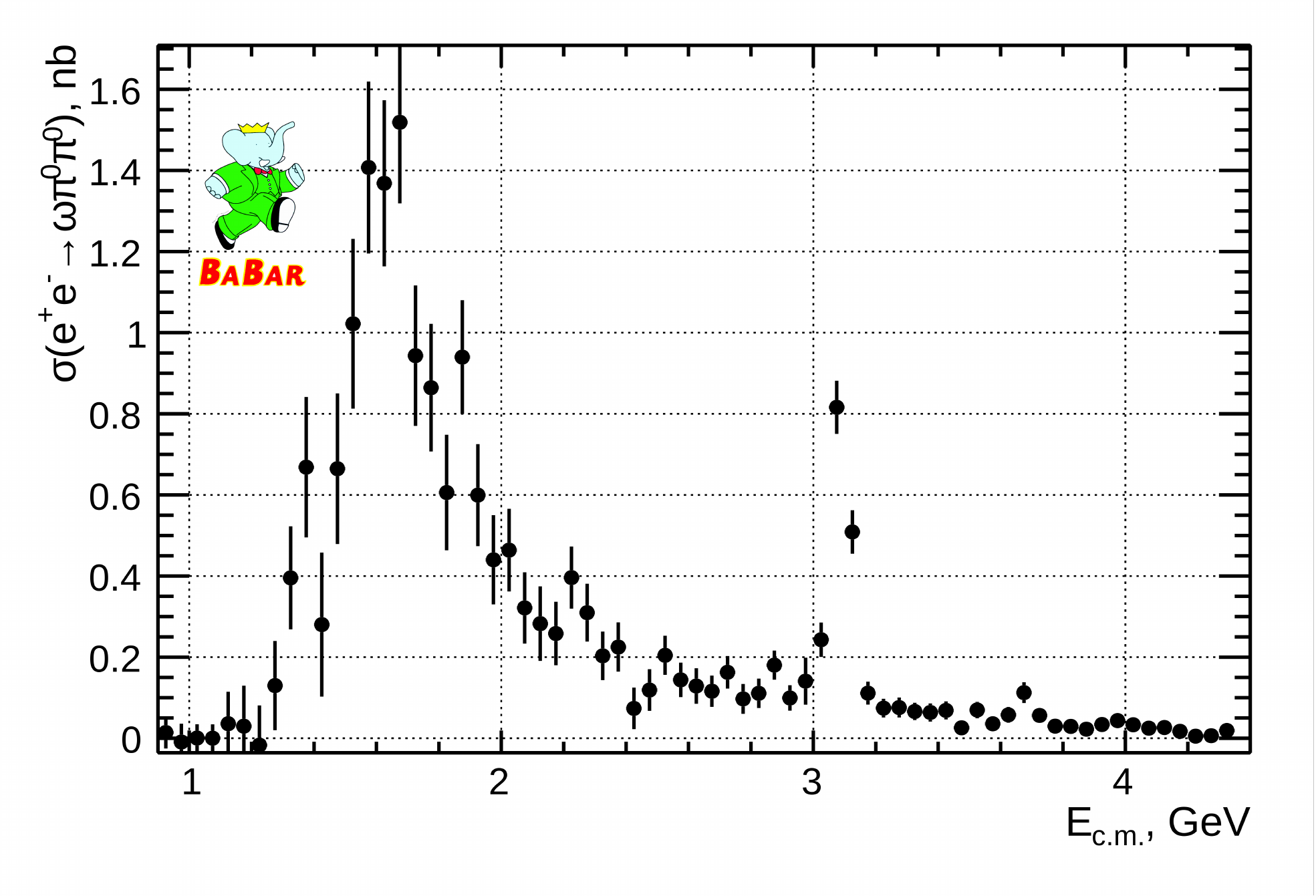}
\caption{Cross-section for $\epem\to\omega\piz\piz$ ISR events in 50\mev intervals in $\sqrt{s^\prime}$ ($E_\mathrm{c.m.^\prime}$).\label{ebert:xsectionomegafig}}
\end{figure}

The $\pip\pim\eta$ and $\omega\piz\piz$ final states are resonant sub-states for the already shown final state $\pip\pim 3\piz$ and therefore included in the result of the 
previous subsection.
The number of events in each energy interval for both final states is shown in fig.\ref{ebert:etaeventsfig} and fig.\ref{ebert:omegaeventsfig}. The resulting 
cross-section distributions over the whole accessible energy range in fig.\ref{ebert:xsectionetafig} and fig.\ref{ebert:xsectionomegafig}.



In addition to those two sub-modes, also finale states with $\rho$ contribute to $\epem\to\pip\pim 3\piz$. Contributions from the final states $\rho^\pm\pi^\mp 2\piz$
and $\rho^+\rho^-\piz$ have been found but are not shown in this conference proceeding for space reasons.

\begin{figure}
\includegraphics[width=0.9\linewidth]{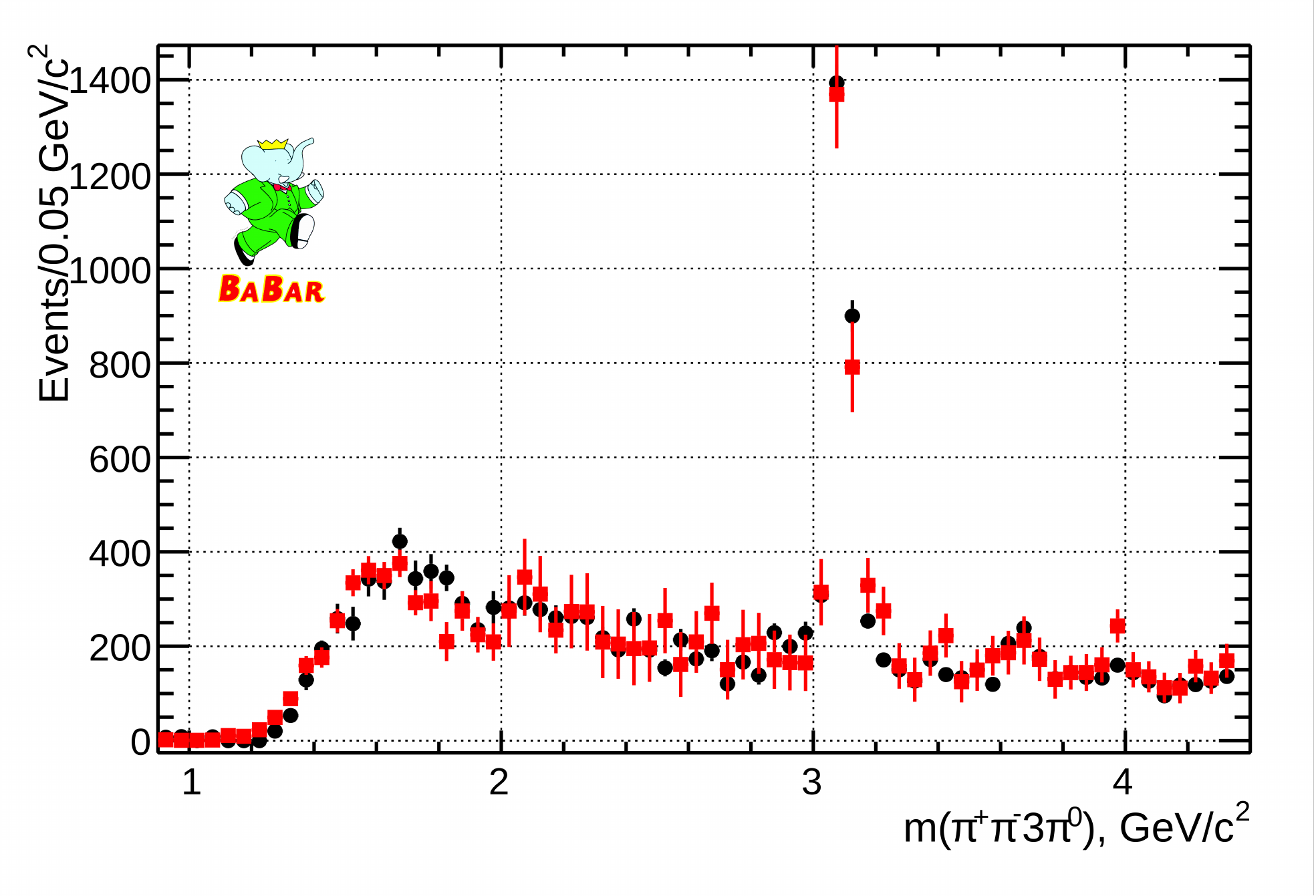}
\caption{Invariant mass distribution $m(\pip\pim 3\piz)$ for $\epem\to\pip\pim 3\piz$ ISR events (black) and for the sum of all resonant sub-modes (red). \label{ebert:eventstotalsumfig}}
\end{figure}

Interesting to note is that the sum of all resonant contributions account for the total number of events found for the final state
$\pip\pim 3\piz$. This is shown in fig.\ref{ebert:eventstotalsumfig}. Therefore it appears that only resonant sub-modes contribute to the $\epem\to\pip\pim 3\piz$ in the studied
energy region.

\subsection{$\pmb{\omega\piz\eta}$ and $\pmb{\phi\piz\eta}$ final states}
\begin{figure}
\includegraphics[width=0.8\linewidth]{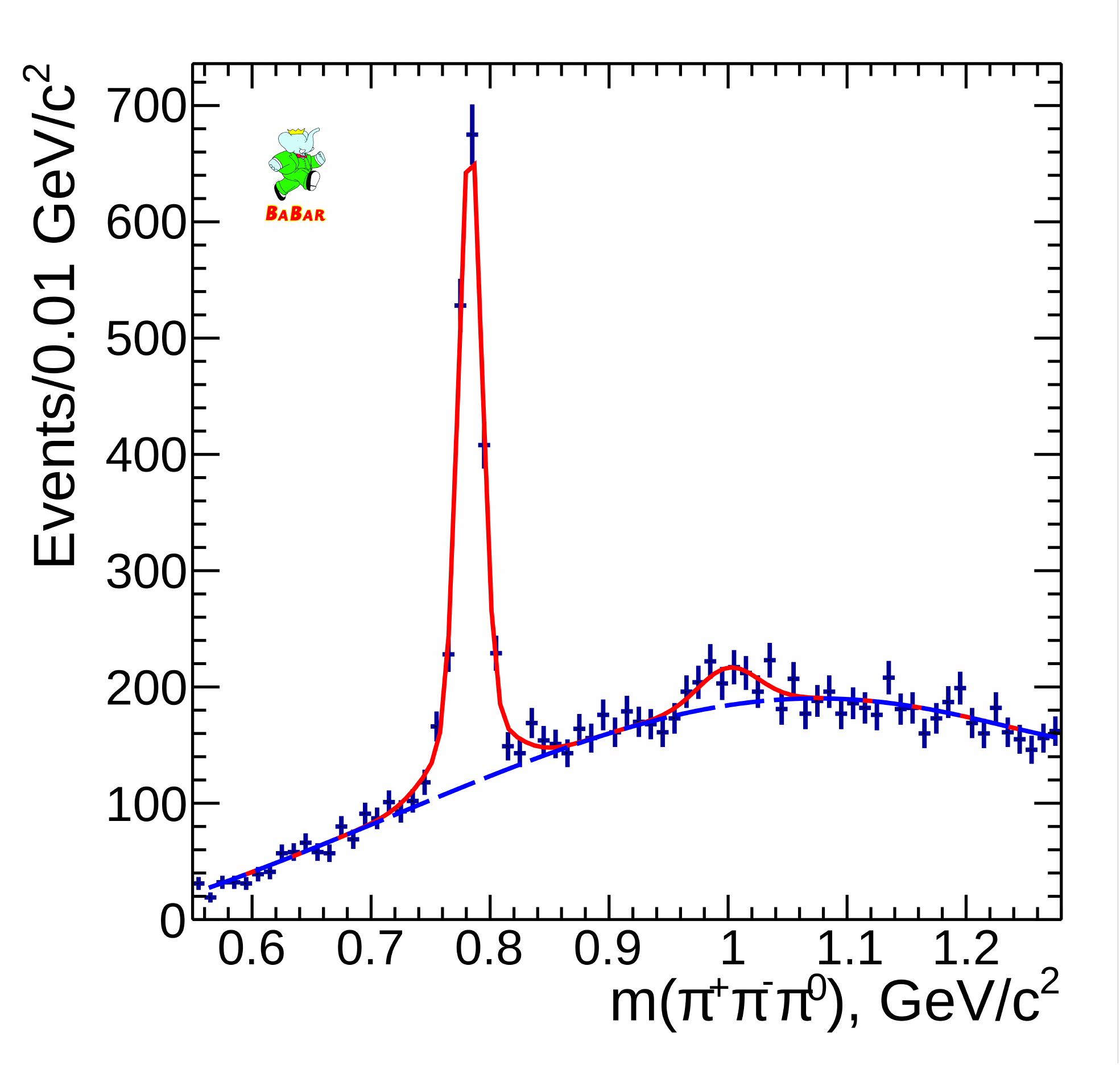}
\caption{Fit to the invariant mass distribution $m(\pip\pim\piz)$ (black) to extract the $\omega$ and $\phi$ yields (red). \label{ebert:eventsomegaphifig}}
\end{figure}
\begin{figure}
\includegraphics[width=0.9\linewidth]{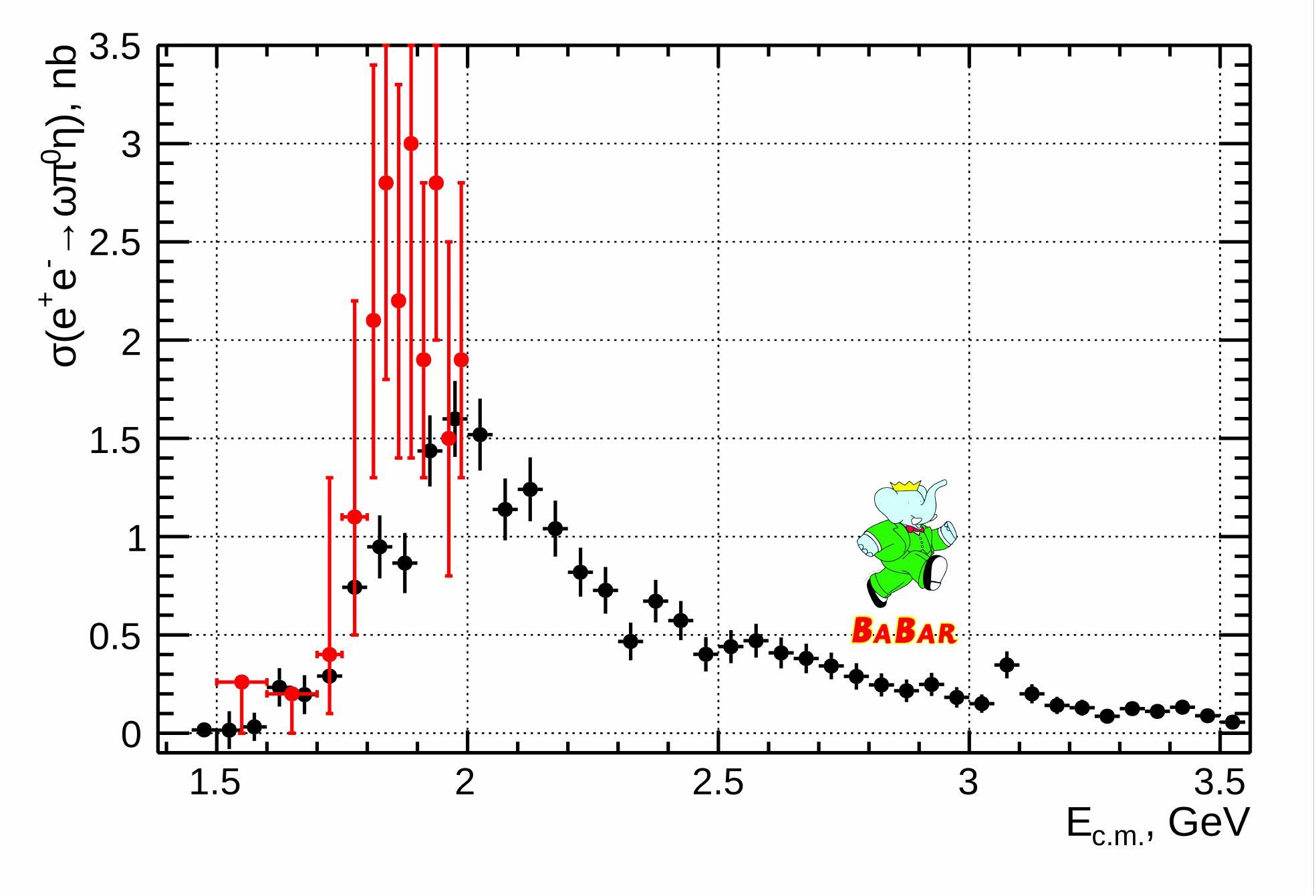}
\caption{Cross-section measurement for $\epem\to\omega\piz\eta$ by \babar~(black) and the SND Collaboration (red). \label{ebert:xsectionomegaphifig}}
\end{figure}
The $\omega\piz\eta$ and $\phi\piz\eta$ final states are resonant sub-states for the previously shown final state $\pip\pim 2\piz\eta$ and include in those measurements.
Figure \ref{ebert:eventsomegaphifig} shows the $m(\pip\pim\piz)$ distribution from which the signal yields for both resonances are extracted by a combined fit.
The cross-section result is shown in fig.\ref{ebert:xsectionomegaphifig} together with the results obtained by the SND collaboration.

\subsection{$\pmb{\jpsi}$ and $\pmb{\psitwos}$ branching fractions}
\begin{figure}
\includegraphics[width=0.8\linewidth]{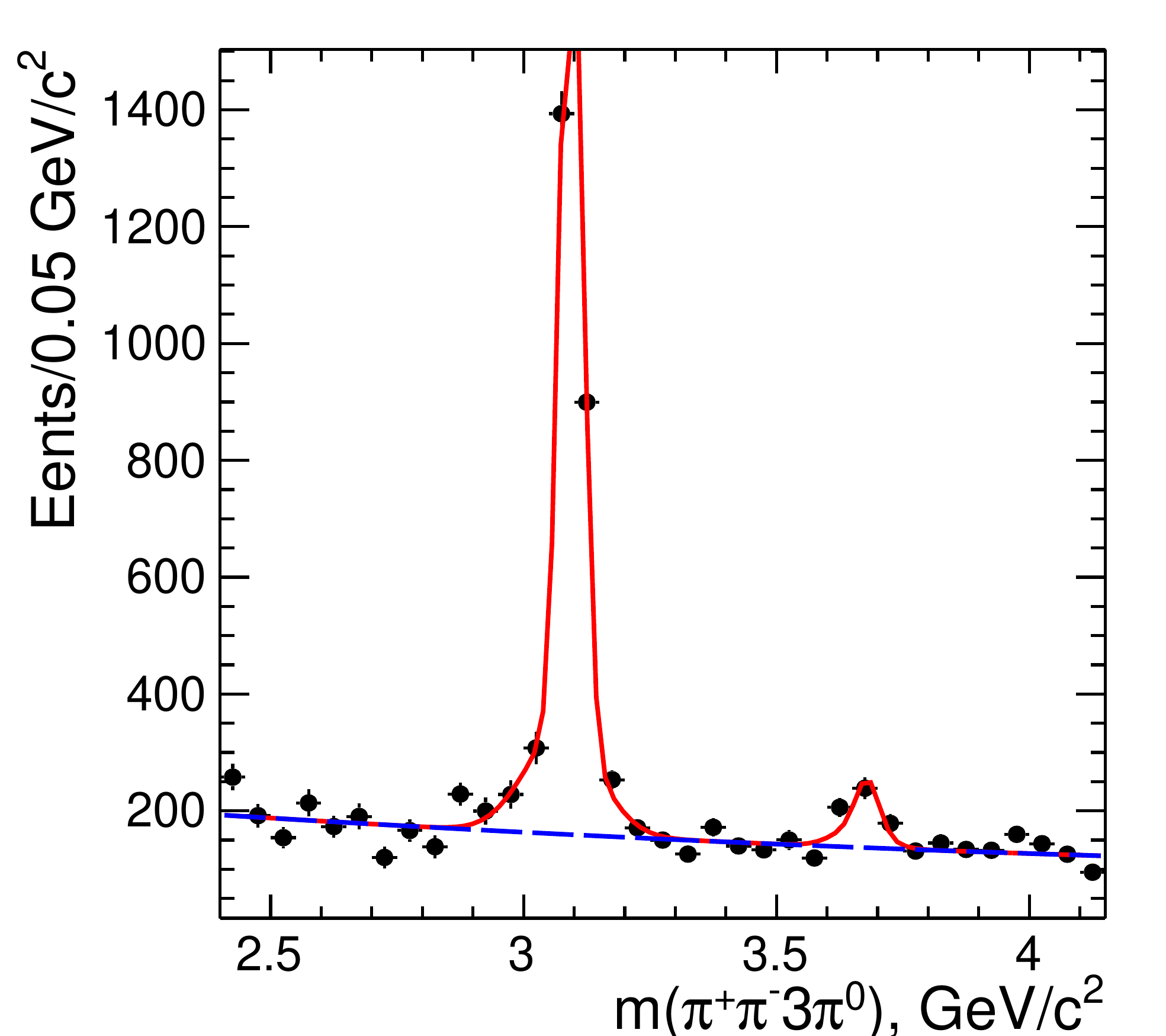}
\caption{Invariant mass distribution $m(\pip\pim 3\piz)$ (black), fitted to extract the number of $\jpsi$ and $\psitwos$ candidates (red).\label{ebert:jpsifig1}}
\end{figure}

\begin{figure}
\includegraphics[width=0.8\linewidth]{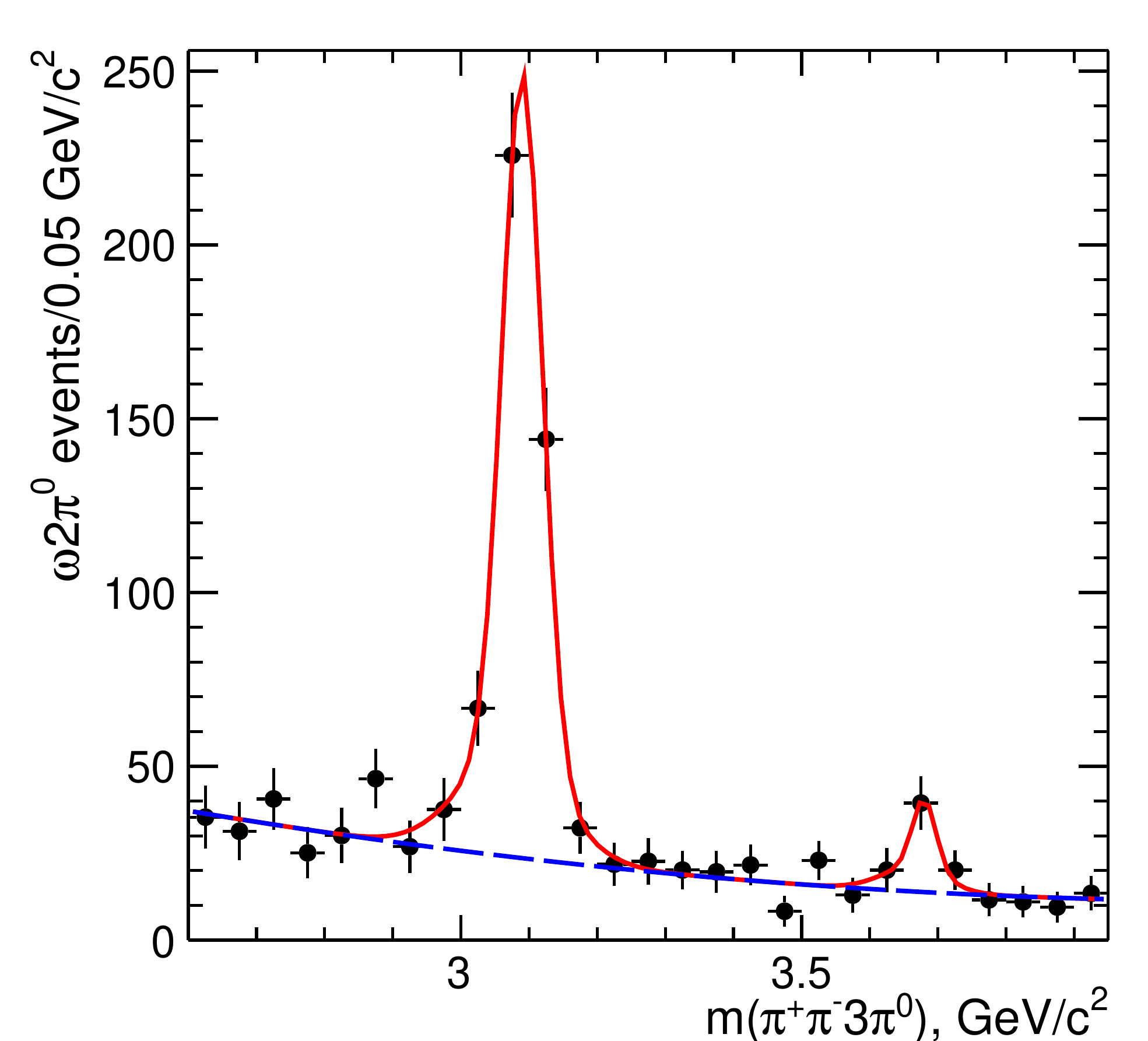}
\caption{Invariant mass distribution $m(\omega 2\piz, \omega\to\pip\pim\piz)$ (black), fitted to extract the number of $\jpsi$ and $\psitwos$ candidates (red). \label{ebert:jpsifig2}}
\end{figure}

\begin{figure}
\includegraphics[width=0.8\linewidth]{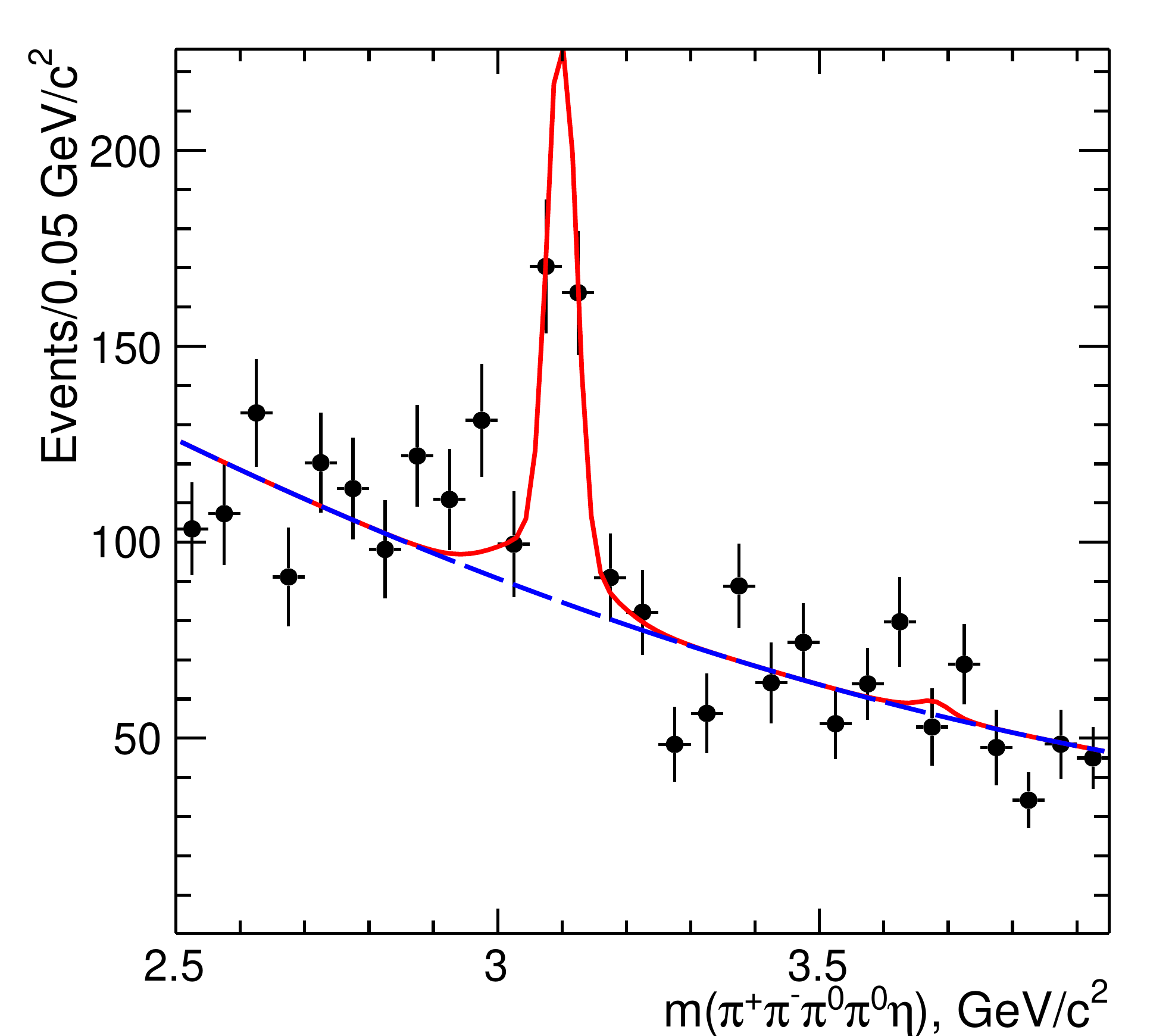}
\caption{Invariant mass distribution $m(\pip\pim 2\piz\eta)$ (black), fitted to extract the number of $\jpsi$ and $\psitwos$ candidates (red). \label{ebert:jpsifig3}}
\end{figure}

When looking to the previously shown distribution of signal events in the different final states, the $\jpsi$ and $\psitwos$ resonances are also seen.
As an example, fig. \ref{ebert:jpsifig1} to \ref{ebert:jpsifig3} show the signal event distribution for different modes around the $\jpsi$ and $\psitwos$ mass.
By fitting those distributions for the number of $\jpsi$ or $\psitwos$ candidates, one can determine the branching fraction for different decay modes of those resonances.
This was done for 12 different modes, shown in table \ref{ebert:bftab}, while only two of them were known before.

\begin{table*}
\begin{tabular}{r@{$\cdot$}l  
 r@{.}l@{$\pm$}l@{$\pm$}l
 r@{.}l@{$\pm$}l@{$\pm$}l
 r@{.}l@{$\pm$}l }
 \multicolumn{2}{c}{Measured} & \multicolumn{4}{c}{Measured}    &
 \multicolumn{7}{c}{$J/\psi$ or $\psi(2S)$ Branching Fraction  (10$^{-3}$)}\\
 \multicolumn{2}{c}{Quantity} & \multicolumn{4}{c}{Value (\ev)} &
 \multicolumn{4}{c}{this analysis \cite{ebert:prdpaper}}    &
 \multicolumn{3}{c}{previously known \cite{ebert:PDG}} \\
 \hline
 $\Gamma^{J/\psi}_{ee}$  &  $\BR_{J/\psi \to \pipi\ppz\piz}$  &
 150& 0 & 4.0 & 15.0  &   ~~~27&0 & 0.7 & 2.7  &   \multicolumn{3}{c}{no entry}  \\
 $\Gamma^{J/\psi}_{ee}$  &  $\BR_{J/\psi  \to \omega\ppz}
 \cdot \BR_{\omega    \to 3\pi}$  &
 24&8 & 1.8 & 2.5  &   5&04 & 0.37 & 0.50  &   ~~~~3&4 & 0.8  \\
 $\Gamma^{J/\psi}_{ee}$  &  $\BR_{J/\psi  \to  \rho^{\pm}\pi^{\mp}\ppz}$&
 78&0 & 9.0 & 8.0  &   14&0 & 1.2 & 1.4  &  \multicolumn{3}{c}{no entry} \\
 $\Gamma^{J/\psi}_{ee}$  &  $\BR_{J/\psi  \to \rho^+\rho^-\piz}  $  &
 33&0& 5.0& 3.3 &   6&0 & 0.9 & 0.6  &    \multicolumn{3}{c}{no entry}\\
 $\Gamma^{J/\psi}_{ee}$  &  $\BR_{J/\psi  \to \pipi\ppz\eta} $ &
 12&8& 1.8& 2.0 &   2&30 & 0.33 & 0.35  &  \multicolumn{3}{c}{no entry} \\
 $\Gamma^{J/\psi}_{ee}$  &  $\BR_{J/\psi  \to \omega\piz\eta}
 \cdot \BR_{\omega\to 3\pi} $  &
 1&7& 0.8 & 0.3 &   0&34 & 0.16 & 0.06  &  \multicolumn{3}{c}{no entry}\\
 $\Gamma^{J/\psi}_{ee}$  &  $\BR_{J/\psi  \to \rho^{\pm}\pi^{\mp}\piz\eta}$ &
 10&5& 4.1 & 1.6 &   1&7 & 0.7 & 0.3  &  \multicolumn{3}{c}{no entry}\\
 $\Gamma^{\psi(2S)}_{ee}$  &  $\BR_{\psi(2S) \to\pipi\ppz\piz} $  &
 12&4& 1.8& 1.2 &   5&2  & 0.8  & 0.5   &  \multicolumn{3}{c}{no entry} \\
 $\Gamma^{\psi(2S)}_{ee}$  &  $\BR_{\psi(2S) \to J/\psi\ppz}
 \cdot \BR_{J/\psi \to 3\pi}  $  &
 10&1& 1.5& 1.1 &   22&9  & 2.8  & 2.3   &   ~~~~~~21&1 & 0.7 \\
 $\Gamma^{\psi(2S)}_{ee}$  &  $\BR_{\psi(2S) \to \omega\ppz}
 \cdot \BR_{\omega     \to 3\pi} $  &
 2&3& 0.7& 0.2 &   1&1 & 0.3 & 0.1  &   \multicolumn{3}{c}{no  entry}\\
 $\Gamma^{\psi(2S)}_{ee}$  &  $\BR_{\psi(2S) \to \rho^{\pm}\pi^{\mp}\ppz} $  &
 ~~~~~~~~$<$6&\multicolumn{3}{l}{2 at 90\% C.L.} &  $<$2  &
\multicolumn{3}{l}{6
at 90\% C.L.}   &    \multicolumn{3}{c}{no entry} \\
$\Gamma^{\psi(2S)}_{ee}$  &  $\BR_{\psi(2S) \to \pipi\ppz\eta} $  &
~~~~$<$0 & \multicolumn{3}{l}{85 at 90\% C.L.} & ~~~~ $<$0 &
\multicolumn{3}{l}{35
at 90\% C.L.} &      \multicolumn{3}{c}{no entry} \\
\hline
\end{tabular}
\caption{Summary of the measured $\jpsi$ and $\psitwos$ branching fractions.\label{ebert:bftab}}
\end{table*}

\section{Conclusion}
Although the \babar~collaboration stopped data taking 11 years ago, it still has a rich ongoing physics program and publishes results of which some are even
first time measurements.
In this presentation, hadronic cross-section distributions were presented for different final states obtained from ISR events over a large energy region. 
These new measurements show 
an increased precision over previous measurements, where previous measurement exist. The presented measurement for the 
cross-section of the final state $\pip\pim 2\piz\eta$ was a first time measurement for this final state. 

In addition, twelve branching fraction measurements for different decay modes of the $\jpsi$ and $\psitwos$ resonances were presented. Only two of them were previously known.

The analyses that measured those cross-sections and branching fractions is published in Phys.~Rev.~D. More details about it as well as tables with all measured
cross-sections can be found in \cite{ebert:prdpaper}.

\bigskip 

\end{document}